\documentclass[usenatbib]{jaa}

\usepackage{times}  
\usepackage{listings}
\usepackage{float}
\usepackage{graphics}
\usepackage{subfigure}
\usepackage{graphicx}
\usepackage{amssymb}
\usepackage{hyperref}
\usepackage{multicol,xcolor}

\newcommand{\aap}{A\&A}
\newcommand{\apj}{ApJ}
\newcommand{\apjl}{ApJL}
\newcommand{\aj}{AJ}
\newcommand{\araa}{ARA\&A}
\newcommand{\apjs}{ApJS}
\newcommand{\mnras}{MNRAS}
\begin{document}

\title{Interactions of galaxies outside clusters and massive groups}

\author{Jaswant K. Yadav\textsuperscript{1} \and Xuelei Chen\textsuperscript{2,3}}
\affilOne{\textsuperscript{1}Department of Physics, Central University of Haryana, Mahendergarh, Haryana 123031, India.\\}
\affilTwo{\textsuperscript{2}National Astronomical Observatories, Chinese Academy of Sciences, 20A Datun Road, Chaoyang District,\\ Beijing $100012$, China\\}
\affilThree{\textsuperscript{3}Center of High Energy Physics, Peking University, Beijing 100871, China }

\twocolumn[{

\maketitle

\corres{jaswantkyadav@cuh.ac.in}


\begin{abstract}
  We investigate the dependence of {physical properties of galaxies on small and large scale density environment}. The galaxy population consists of mainly passively evolving galaxies in comparatively low density regions of Sloan Digital Sky Survey (SDSS).  We adopt  (i) local density, $\rho_{20}$, derived using adaptive smoothing kernel, (ii) projected distance, $r_p$, to the nearest neighbor galaxy and (iii) the morphology of the nearest neighbor galaxy as various definitions of environment parameters of every galaxy in our sample. In order to detect long-range interaction effects we divide galaxy interactions into four cases depending on morphology of target and neighbor galaxies.  {This study builds upon an earlier study by Park \& Choi (2009) by including { improved definitions} of target and neighbor galaxies thus enabling us to better understand the effect of ``the nearest neighbor'' interaction on target galaxy}.    
We report that the impact of interaction on galaxy properties is detectable at least out to the pair separation corresponding to the virial radius of (the neighbor) galaxies. This turn out to be mostly between 210 and 360 $h^{-1}$kpc for galaxies included in  our study.
We report that early type fraction, for isolated galaxies with $r_p > r_{vir,nei}$ are almost ignorant of the background density and, has a very weak density dependence for closed pairs. Star formation activity of a a galaxy is found to be crucially dependent on neighbor galaxy morphology. We find star formation activity parameters and structure parameters of galaxies to be independent of the large scale background density. We also exhibit that changing the absolute magnitude of the neighbor galaxies does not affect significantly the star formation activity of those target galaxies whose morphology and luminosities are fixed.
\end{abstract}

\keywords{galaxies: evolution -- galaxies: interactions -- galaxies: fundamental parameters -- galaxies: general}
}]

\year{2018}
\pgrange{1--14}
\setcounter{page}{1}
\lp{14}

\label{firstpage}

\section{Introduction}

One of the key feature of the hierarchical picture of galaxy 
formation and growth involves continuous interactions and mergers
with other galaxies. Various galaxy properties such as morphology, luminosity, structure parameters, star-formation
rate (SFR) and  dust properties are strongly affected by these interactions ( Struck 2006; Woods \& Geller 2007; Blanton \& Moustakas 2009).
The location of galaxies in particular environment is known to significantly affect their physical properties (Hubble 1936; Oemler 1974; Dressler 1980; Postman \& Geller 1984; Kauffmann {\em et al.}
2004; Blanton {\em et al.} 2005b; Weinmann {\em et al.} 2006; Blanton \& Berlind 2007; Lambas {\em et al.} 2012). With the advent of large spectroscopic and photometric surveys, the study of the role of the environment in galaxy formation has got a new impetus (Balogh {\em et al.} 2004; Park {\em et al.} 2007;
Cooper {\em et al.} 2006; Cucciati {\em et al.} 2006; Skibba {\em et al.} 2012).  With the availability of large galaxy samples, such as Sloan Digital
Sky Survey Data Release 7 [SDSS-DR7] (Abazajian {\em et al.} 2009), we are in a position to explore many dimensions of galaxy properties simultaneously and homogeneously. This kind of analysis helps us  put galaxy scaling relationships in context with respect to one another as well as probe the underlying physics affecting galaxy formation and evolution.

Starting from the suggestions of Toomre (1977) that elliptical galaxies can be formed by the merger between spiral galaxies, there is growing evidence for a change in galaxy morphology (e.g. Park {\em et al.} 2008; Buta 2013) and galaxy structure as a result of merger (e.g. Nikolic {\em et al.}
2004; Patton {\em et al.} 2005; HernÃ¡ndez-Toledo {\em et al.} 2005; Park \& Choi 2009). On the other hand, at fixed color, the residual dependence of galaxy morphology on environmental density is reported to be rather weak (Ball {\em et al.} 2008; Bamford {\em et al.} 2008).  With an earlier release of SDSS data (DR4) Park {\em et al.} (2008) showed that when a galaxy is located within the virial radius of its nearest neighbor, its morphology tends to be the same as that of the neighbor. This points to an important role of hydrodynamic interactions with neighbors within the virial radius. Further studies comprising of galaxies in galaxy clusters (Park \& Hwang 2009; Cervantes-Sodi {\em et al.} 2011), and involving hosts and their satellite galaxies (Ann {\em et al.} 2008; Wang {\em et al.} 2010) seem to favor this proposition.

In regard to star formation activity, following Zwicky's suggestion that collisions would be frequent within dense galaxy clusters,  Spitzer \& Baade (1951) argued  that strong shock waves could push the interstellar gas out of these galaxies resulting in scarcity of late-type spiral galaxies with substantial ongoing star formation in clusters. Larson \& Tinsley (1978) identified interacting galaxies from Atlas of Peculiar Galaxy, and determined that many of these galaxies had recently undergone a burst of star formation  (see also McQuinn {\em et al.} 2010).
Several groups (e.g. Condon {\em et al.} 1982; Keel {\em et al.} 1985; Kennicutt {\em et al.} 1987) have observed and found bursts of star formation associated with tidal interactions. Barton {\em et al.} (2007) constructed an isolated ``field'' galaxy mock sample using cosmological N-Body simulations and compared it with ``field'' galaxy sample from 2dF data to establish that a large fraction of the galaxies that experience close passes respond with triggered star formation. Tonnesen \& Cen (2012) find the tendency of  close pairs in high-density environments to have fewer low specific SFR galaxies than non-pairs, whereas pairs in low-density environments have the opposite trend. Star formation related properties, such as color and emission-line flux, are found to be directly correlated with environmental density (Kauffmann {\em et al.} 2004; Blanton {\em et al.} 2005b; Christlein \& Zabludoff 2005; Patton {\em et al.}
2011; Ideue {\em et al.} 2012).  In addition, SFR and the  fraction of star-forming galaxies is reported to be decreasing with increasing galaxy density (Carter {\em et al.} 2001; Lewis {\em et al.} 2002; G{\'o}mez {\em et al.}
2003; Mateus \& SodrÃ© 2004; Mahajan {\em et al.} 2010; Ellison {\em et al.} 2011. 

Regarding studies of structure parameters of galaxies, Bretherton {\em et al.} (2013) studied the normalized rates and radial distributions of star formation in galaxies within low-redshift clusters to report that morphological type classifications of cluster galaxies  correlate only weakly with their concentration indices, whereas this correlation is strong for non-cluster populations of disk galaxies. Park \& Choi (2009) studied the variation of velocity dispersion and found almost no change for early type galaxies while for late type galaxies the velocity dispersion becomes larger as they approach their neighbors. 
Huertas-Company {\em et al.} (2013) report that at intermediate redshifts, the  size-mass relation for passive early-type galaxies is independent of environments ranging from field to groups which seem to contradict the findings by Cooper {\em et al.} (2012)  at similar redshifts. Lani {\em et al.} (2013) find that the  massive quiescent galaxies at $z > 1$ are typically 50\% larger in the highest density environments compared to those in the lowest density environments whereas this relationship between size and environment is much weaker for star-forming galaxies

In addition to small scale dependence of galaxy properties discussed above, there are different studies which  determine the dependence of various galaxy properties on large-scale structure (Binggeli 1982; White {\em et al.} 2010). Einasto {\em et al.} (2007) reported  higher fraction of early type, passive, red galaxies in rich superclusters compared to poor superclusters thereby indicating the role of global environment in influencing galaxy morphology and their star formation activity (see
also Patel {\em et al.} 2011; Lietzen {\em et al.} 2012). Park {\em et al.} (2007), however, concluded that different galaxy properties are independent of large-scale density once morphology and luminosity are fixed (see also Park \& Choi 2009; Kajisawa {\em et al.} 2013).

 The extent to which the galaxy properties are driven by internal-physical-processes ('\textit{nature}') as opposed to external-physical-processes ('\textit{nurture}') is still a matter of debate (Harrison {\em et al.} 2011).  A sample of galaxies whose  physical properties are largely the result of internal-physical-processes can be used to address this issue in greater details.  By comparing the physical properties of these galaxies in different environments we can have a better idea of the role played by  nurture-induced processes. Keeping this in mind, for this work, we use a subsample of galaxies from SDSS-DR7 which are neither in the vicinity of a member of Abell Cluster of galaxies nor a part of large group of galaxies. The three-dimensional environment parameter space, used in this study, enables us better understand the effects of galaxy interactions.

A brief outline of the paper follows. In section \ref{data} we describe the data from Sloan Digital Sky Survey. Different subsections describe associated morphology classification. small and large scale density environments, definition of nearest neighbor galaxy as used in our analysis as well as dataset comparison to previous studies in the literature. The results of environment dependence of different galaxy properties are presented in detail in section \ref{result}. We conclude this paper with discussions and conclusions in section \ref{discuss} 

\section{Data} \label{data}
\subsection{SDSS Galaxy Sample}
We use galaxies from the SDSS-DR7. This data release covers about 8,032 square degrees in the northern sky and consists of a series of three interlocking imaging and spectroscopic surveys. A $2.5$m telescope located at Apache Point Observatory in Southern New Mexico is being used to carry out observations in five photometric bands ranging in wavelength from $3000$ to $11000$ $\AA$ (Fukugita {\em et al.} 1996; Gunn {\em et al.} 1998). After careful photometric image reduction, calibration and classification of individual galaxies, a subsample of galaxies with   Petrosian $r$ band magnitude $r < 17.77$ has been chosen for follow up spectroscopic observations (Lupton {\em et al.} 1999, 2001; Strauss {\em et al.} 2002). The spectroscopic sky survey is being  performed using two multi-object fiber spectrographs on the same telescope. Each spectroscopic fiber plug plate having a circular field-of-view, with a radius of 1.49 degrees, can accommodate a total of 640 fibers. Because of the finite size of the fiber plugs, the minimum separation of fiber centers is $55''$. If two galaxies are within $55''$ of each other, both of them can be observed only if they lie in the overlap between two adjacent fiber plug plates (Blanton {\em et al.} 2003).

\begin{figure}[htb]
\begin{center}
\includegraphics[height=11cm,width=9cm]{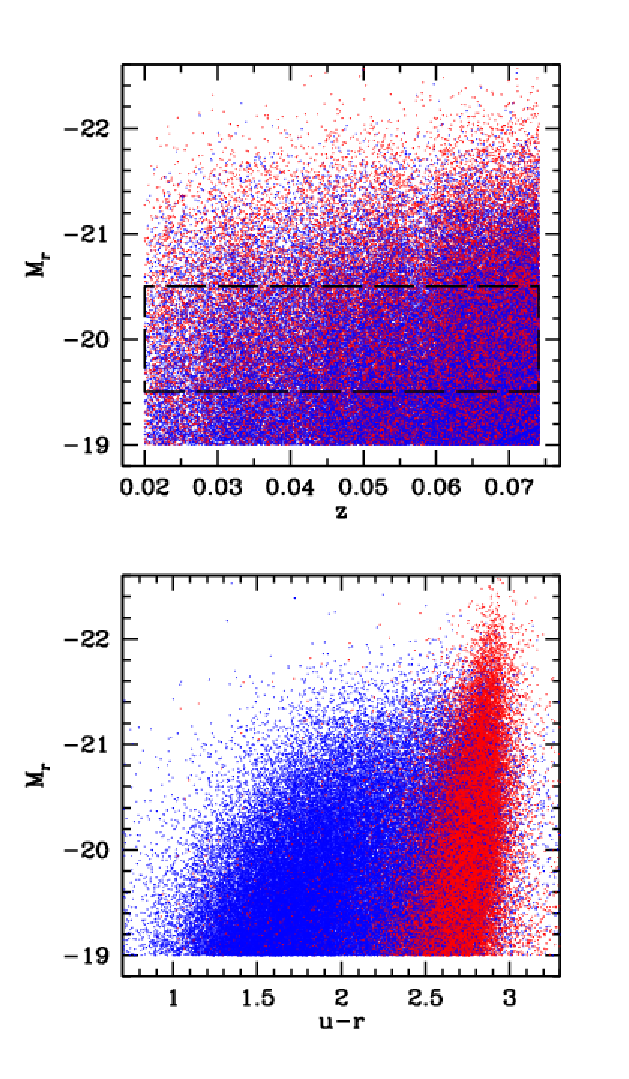}
\end{center}
\caption{The rectangular box in the upper panel encloses the target galaxies which are part of a total of $114,272$ galaxies contained in our volume limited subsample. The faint limit of these target galaxies is $0.5$ mag brighter than the full sample to achieve complete neighbor selection. The bottom panel shows the galaxies in the $color-magnitude$ diagram. Red points are early$-$morphological$-$type galaxies and blue points are late types.}\label{fig1}
\end{figure}

For the present study we use the Korea Institute for Advanced Study Value-Added Galaxy Catalog (KIAS-VAGC) (Choi {\em et al.} 2010) This catalog has been extracted from the New York University Value-Added Galaxy Catalog (NYU-VAGC) Large Scale Structure Sample (brvoid0) (Blanton {\em et al.} 2005) that includes 583,946 galaxies within a $r$ band apparent magnitude range of $10 <  r <17.6$. Our sample is supplemented by brighter galaxies that are not part of SDSS and whose redshifts are obtained from various literatures (Choi {\em et al.} 2007). This leads to a total of $707,817$ galaxies in the flux limited sample. Such a large catalog of galaxies offers the advantage of an extended magnitude range with high completeness. The volume limited sample used for the present study is a galaxy sample with $r-$band absolute magnitude $M_r < -19.0 +5~log~h$ and redshifts $ 0.020 <$ z $< 0.074$ or a comoving distance of $59.7$ $h^{-1}$Mpc $<$ d $< 218.9$ $h^{-1}$Mpc. This sample includes $114272$ galaxies. In order to select, mainly passive galaxies from low density environments, we post process the volume limited sample by removing galaxies within a velocity separation of $\pm$ $3000$~kms$^{-1}$ and a projected separation of $3$ Mpc from galaxies located inside Abell Clusters (Abell {\em et al.} 1989) of galaxies. The subsample of galaxies is plotted in figure \ref{fig1}.   

\subsection{Morphology Classification}

Park \& Choi (2009) stressed the importance of accurate morphology classification since the effects of interaction strongly depend on morphology of interacting galaxies. For this work we use Park \& Choi (2005) prescription to classify  morphological types of galaxies in our sample. In this scheme the the location of galaxies, in the color-color gradient space, combined with their i-band concentration enables us in dividing the sample  into early (ellipticals and lenticulars) and late (spirals and irregulars) morphological types with completeness and reliability of sample around 90\%. An additional visual check of color images is undertaken to account for possible incorrect morphological classification of about 10,000 galaxies  having close neighbors. This is necessary because Park \& Choi (2005) prescription performs poorly when an early-type galaxy starts to overlap with other galaxies. Visual analysis  guides us in identifying blue but elliptical-shaped, and dusty edge-on spiral galaxies. It is also useful in correcting central positions of merging galaxies. These procedures result in a volume limited subsample of 114,272 galaxies with $M_r$ $<$ $-−$19.0. Removing the galaxies that are, part of massive SDSS groups (Yang {\em et al.} 2007), in the vicinity of Abell Clusters or are near the survey boundaries result in a population of 
$86,604$ galaxies out of which $30,298$ are early-type, and the rest are late-type galaxies. 

Different properties of galaxy population are studied in the absolute magnitude range of $-19.5 > M_r > -−20.5$. In such a subsample volume we have $14,244$ early types and $25,771$ late types. We identify these galaxies as those lying within a rectangular box in
Figure \ref{fig1}. As argued in literature (see e.g. Choi {\em et al.} (2007); Park \& Choi (2009)), in order to minimize the effect of internal extinction on different galaxy properties, we restrict our late-type galaxy population to those having an isophotal axis ratio greater than $0.6$. This results in a further reduction of late-type galaxies to $14,581$ within the absolute magnitude limit of our subsample. 
Different physical properties of galaxies are studied by subdividing the sample of galaxies into four subsamples: $7106$ early types having early type nearest neighbor (the E$-$e galaxies), $7138$ early types having late-type nearest neighbor (E$-$l), $6894$ late types having early-type nearest neighbor (L$-$e), and $7687$ late types having late-type nearest neighbor (L$-$l). 
\subsection{Environment}
There is a variety of methods to measure galaxy environment that may or may not correlate with each other (Muldrew {\em et al.} 2012). We have followed Park \& Choi (2009) in defining three different environment measures namely large-scale background density, small-scale mass density and morphology of the closest neighbor galaxy. In this scheme large scale background density at given location of an object in our sample is given by 
\begin{equation}
  \rho_{20}({\bf x})/\bar\rho = \sum_{j=1}^{20}\gamma_j L_j W_j(|{\bf x_j - x}|) / \bar\rho
  \label{eq100}
\end{equation}
where $\gamma$ and $L$ are mass to light ratio and luminosity of closest $20$ galaxies around an object in our volume limited sample.  $\bar\rho$ is the mean mass density of the universe obtained from the galaxy catalog of total volume $V$, via the relation $\bar\rho = \sum_{all}^{}\gamma_i L_i/V$, where the summation is over all galaxies brighter than $M_r = -19$. {We report a value of $\bar\rho$ to be $0.02228(\gamma L)_{-20}~h^{-3} Mpc^{-3}$ with $(\gamma L)_{-20}$ being the mass of a late type galaxy with $M_r = -20$}. The weighting, $W$, is defined in terms of a spline kernel (Monaghan \& Lattanzio 1985; Park {\em et al.} 2007) because it is adaptive, centrally weighted and has a finite tail, making it superior to tophat, cylindrical and Gaussian kernels. The smoothing scale determined by $20$ nearest neighbor in our volume limited sample is larger than typical cluster virial radius (about $1$-$2$h$^{-1}$Mpc) resulting into an estimation of $\rho_{20}$ that never exceeds the virialization density. 

The small scale density experienced by a target galaxy due to local mass density given by its nearest neighbor is defined as 
\begin{equation}
\rho_n/\bar\rho = 3\gamma_n L_n/4\pi r_p^3 \bar\rho
\end{equation}
with $r_p$ being the projected separation of the nearest neighbor from target galaxy. Following Park {\em et al.}(2008), we have defined the virial radius of a galaxy as the projected radius where the mean mass density within the sphere of radius $r_p$ is $200$ times the critical density of the universe. According to our estimates, the virial radii of galaxies with $M_r = -19.5, -20.0$ and $-20.5$ are $264.74, 308.66$ and $359.87h^{-1}$kpc for early types and  $210.12,~244.98$ and $285.63h^{-1}$kpc for late types, respectively. Clearly, the virial radius thus defined includes the galaxy as well as the surrounding dark matter halo region.   

\subsection{The Nearest Neighbor} \label{nn}
The definition of nearest neighbor galaxy in our study is slightly modified compared to the one in Park \& Choi (2009). It requires the neighbor to be situated closest to the galaxy on the sky satisfying absolute magnitude and radial velocity conditions as well. The absolute magnitude condition requires the difference between neighbor and target galaxy's absolute magnitude to be smaller than $\Delta M_r$ and larger than $\Delta M_r - 1 $. For most of the work we chose $\Delta M_r$ to be 0.5. We also discuss, in section \ref{discuss}, the dependence of galaxy properties on  a range of  values of $\Delta M_r$. The radial velocity condition demands that the neighbor galaxy must lie within $\pm 600 (800) $kms$^{-1}$ if the target is late(early) type. The velocity difference between target and neighbor galaxy should also depend on their luminosity (Faber \& Jackson 1976; Tully \& Fisher 1977) and projected separation. Accordingly velocity dispersion between early and late-type galaxies in our sample turns out to be between $1.3$ to $1.4$ across different separations(see Figure 2 of Park \& Choi (2009) for details).

\section{Improvements over Earlier Studies} \label{compr}
Our analysis with regard to dependence of galaxy properties on environment builds upon a similar exercise carried out in Park \& Choi (2009) on an earlier release of SDSS data. {The definition of nearest neighbor in our study improves upon the previous definition by introducing a cut in the Absolute magnitude for comparatively brighter neighbor galaxy. This restricts our target-neighbor galaxy sample to lie within a small absolute magnitude range, {resulting in the variation of different galaxy properties purely due to interaction between neighbors and not due to variation of absolute magnitude within the sample}}. Furthermore, in this study, we apply finger of god corrections (Bahcall {\em et al.} 1986) to galaxies belonging to SDSS group catalog of Yang {\em et al.} (2007) and remove galaxies which are member of massive (more than 10 member) groups. {This is done to minimize the effects of large scale density environments on various galaxy properties that were present in Park \& Choi (2009). The properties of a galaxy in our subsample, therefore, are affected purely by the galaxy's own dynamics and not necessarily by external environment provided by hot gas of groups and clusters}. In earlier studies the property of a galaxy could be affected by more than one neighbor galaxy as the target galaxy could lie within the influence region of second and higher nearest neighbor. For the present study a target galaxy  that is  within virial radius of its second nearest neighbor (refer to \ref{nn}) is removed from sample. The removal enables  us to better understand the effect of {``the nearest neighbor''} on different galaxy properties.

\section{Results}\label{result}
We study the variation of galaxy properties on distance to the nearest neighbor, morphology of the nearest neighbor and background density contributed by $20$ neighbor galaxies. We leave out galaxies from very high density environments like Abell clusters and massive SDSS groups. For the galaxies in our subsample, we study different properties such as morphology, absolute magnitude, star formation activity (u-−r color, g-−i color gradient, equivalent width of the H$\alpha$ line) and structure parameters (central velocity dispersion, i-band Petrosian radius, concentration index).
\begin{figure}[htb]
\begin{center}
\includegraphics[height=11cm,width=10cm]{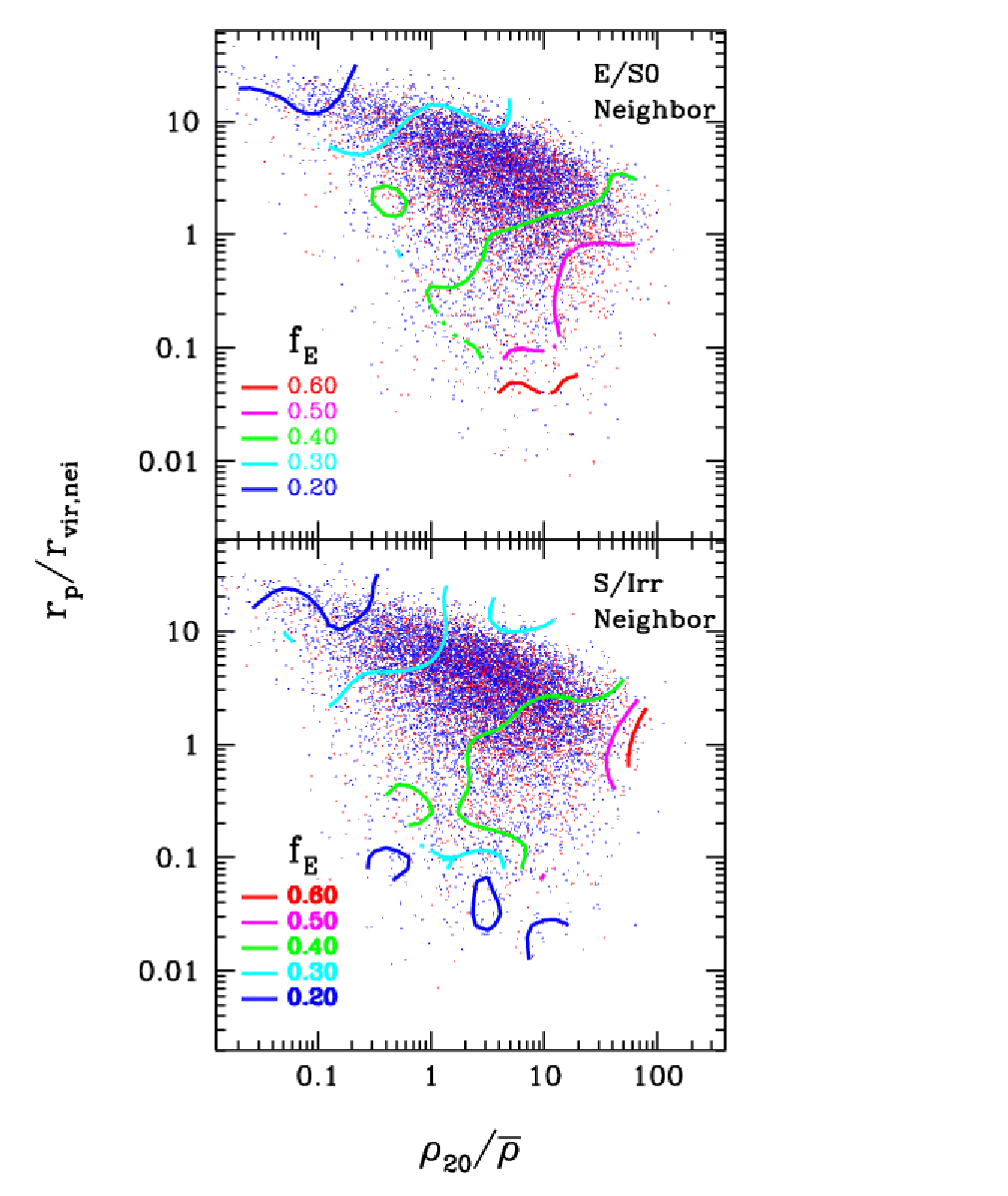}
\end{center}
\caption{(Upper) Environment dependence of Morphology, when the nearest neighbor is an early-type galaxy. Blue (and red) points are late (and early)-type target galaxies. Absolute magnitude of galaxies is confined within $−-19.5 >$ $M_r$ $> -−20.0$. Constant
early-type galaxy fraction, $f_E$, contours are limited to regions with higher than 1$\sigma$ statistical significance. (Lower) Same, but for the  nearest neighbor galaxy being late-type.}\label{fig2}
\end{figure}

For most portions of our study, we fix the r-band absolute magnitude of target galaxies in a narrow range between $-19.5$ and $-−20.5$. This enables restricting effects due to the coupling of a parameter with luminosity.  As pointed out in Choi {\em et al.} (2007), late-type target  galaxies with the i-band isophotal axis ratio less than $0.6$ suffer from internal extinction and corresponding dispersion in luminosity. To avoid false trends in galaxy properties due to this issue, we have discarded such galaxies from our analysis.

\begin{figure*}[ht]
\begin{center}
\includegraphics[height=12cm,width=13cm]{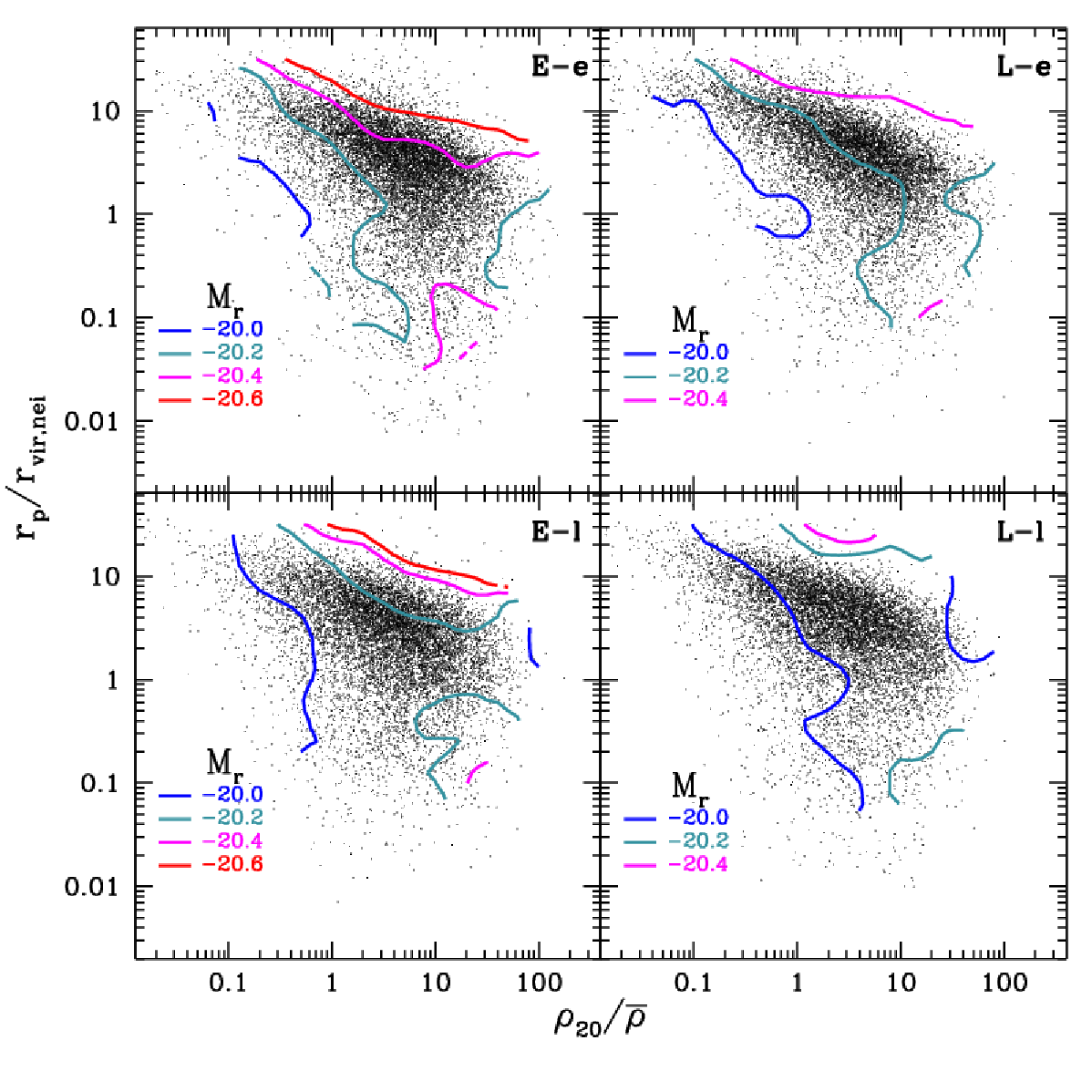}
\end{center}
\caption{Three dimensional (morphology, $r_p$ and $\rho_{20}$) environment dependence of Median absolute magnitude. Points represent target galaxies more luminous than $M_r$ = $-19.5$. The early-type target galaxies with an early-type neighbor (E-e), early-type targets with a late-type neighbor (E-l), late-type targets with an early-type neighbor (L-e) and  late-type targets with a late-type neighbor (L-l) are represented in four panels.}\label{fig4}
\end{figure*}

Smooth distributions of various physical parameters in figure \ref{fig4}, \ref{fig6} and \ref{fig8} are found by the following method. At each point of parameter space, we first sort the values of physical parameter of galaxies contained within a certain spline radius from the point. The median value of a physical parameter is then given by the galaxy whose cumulative sum of spline kernel weights is $\sum w_i/2$, where $w_i$ is the individual spline weight of a galaxy around that point. The median  is preferred over mean because of possible skewness in the distribution of various physical properties. 
 
\subsection{\it Morphology}
We build our study on previous similar analysis (e.g. Park \& Choi 2009; Park \& Hwang 2009) of morphology dependence on local environment. With a much larger SDSS DR7 data sets we are in a position to study this dependence in greater details for galaxies lying in comparatively low density regimes. Figure \ref{fig2} shows the variation of fraction $f_E$ of early type galaxies as function of large scale background density ($\rho_{20}$) and projected separation $r_p$ from the galaxy's nearest neighbor. Early (red dot) and late (blue dot) type galaxies, having early type (upper panel)  and late type (lower panel) nearest neighbor, are distributed in a triangular shaped region in this figure. This is due to the statistical correlation between $r_p$ and $\rho_{20}$.
At each point of parameter space we obtain the value of $f_E$ from the ratio of weighted number of early type galaxies to the weighted number of total galaxies within a smoothing spline kernel of fixed size.  The contours thus obtained are restricted to regions with statistical significance above  1$\sigma$.  {In case of target galaxies having early type neighbors (upper panel), we detect a gradual increase in $f_E$ as the galaxy approaches its neighbor. For galaxies in our sample we find no (or very little) dependence of $f_E$ on background density for galaxies that are situated outside (or inside) the virial radius of their neighbors, thus indicating the absence of any  morphology density relation (Dressler 1980)}. The behavior of $f_E$ for galaxies with late type nearest neighbor (lower panel) is similar at separations larger than neighbor's virial radius. However, when the galaxy is inside the virial radius of its neighbor, $f_E$ starts decreasing after attaining a maximum value around $r_p$ $\sim$ 0.3 $r_{vir,nei}$. {Significant hydrodynamic effects from neighbor galaxies appear to be important at these separations}. Sensitivity of $f_E$ to $\rho_{20}$ exists mainly within the virialized region at very high background density.  This verifies earlier studies (e.g. Park {\em et al.} 2008) confirming that  the effects of the nearest neighbors are critically important to galaxy morphology. {In the region of low merger and interaction ($r_p > r_{vir,nei}$), we report an early type fraction that asymptotically approaches about 0.2}. This could be the inborn morphology fraction. In such an environment, early type galaxies couldn't have formed by processes such as strangulation because of lack of nearby large galaxies.  According to recent observational (Croton \& Farrar
2008) and theoretical (Hirschmann {\em et al.} 2013) studies, {such early type galaxies could either have formed due to AGN heating or other internal processes of galaxy formation}.  
\subsection{\it Luminosity}

Figure \ref{fig4} examines the environmental dependence of $M_r$ in the $r_p - \rho_{20}$ plane. 
We find that on smaller separation from the neighbor, i.e. at $r_p/r_{vir,nei} <1 $, the galaxies  irrespective of their morphology show no variation in $M_r$ as a function of separation and almost none to very little ($\sim 0.2$) variation with respect to the underlying density field. At much smaller separation, we also observe a tendency for a galaxy to have neighbor of same morphology, which is indicated from a lesser population of $E-l$ galaxies compared to $E-e$ galaxies. Park {\em et al.} (2008, 2009) have earlier reported an early to late type morphology transformation which we can explain 
by larger number of $L-l$ galaxies in comparison to $E-l$ galaxies at the scale of hydrodynamic interactions. { Such a transformation can be explained by transfer of cold gas from late type neighbors to their early type target galaxies through interactions, thereby changing the morphology of early type galaxies to late type}. However when the galaxies are well separated from their neighbor, we find significant dependence on both separation as well as density field. In such a situation, we find that galaxies that are farther from their neighbors tend to be the brighter one. The luminosity decreases as the galaxies come closer at this separation and this behavior is more apparent in high density environments. In the low density environments the variation with respect to $r_p/r_{vir,nei}$ is slow possibly because of fewer number of neighbors than Park \& Choi (2009). Another observation from figure \ref{fig4}  helps us to infer that {higher luminosity contours have negative slopes, indicating that the change in $M_r$ due to $\rho_{20}$ is much faster when the galaxies are at comparatively larger separation}.

\subsection{\it Star Formation Activity (SFA) Parameters}
\begin{figure*}[htb]
\begin{center}
\includegraphics[height=18cm,width=19cm]{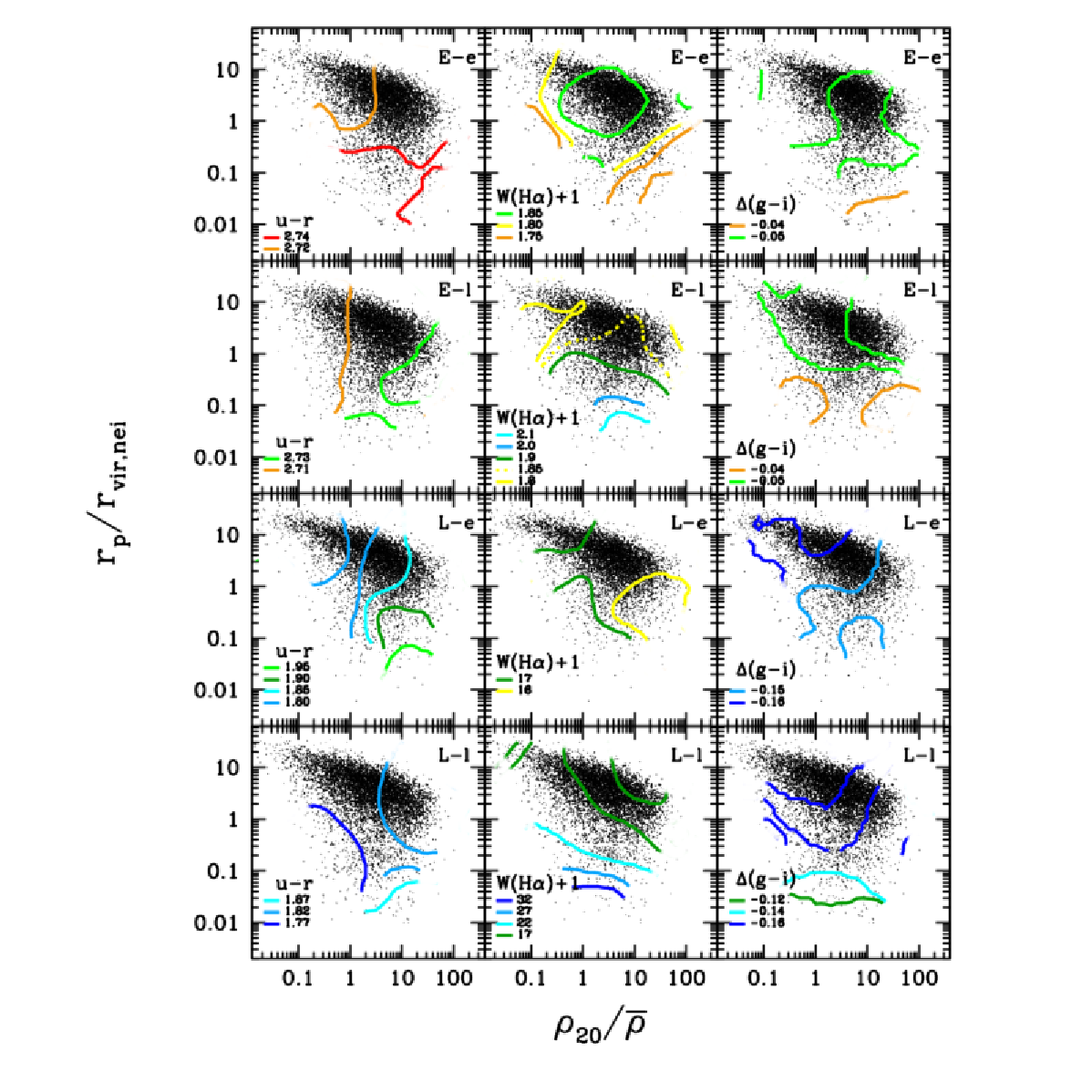}
\end{center}
\caption{
Variation of $u-−r$ color, equivalent width of the the $H\alpha$ line and $g-−i$ color gradient of galaxies with $-−19.5 > Mr > -−20.5$ with respect to  the pair separation $r_p$ and the large-scale background density $\rho_{20}$. In every column, target galaxies are divided into four cases: the E-e, E-l, L-e, and L-l galaxies, respectively. {At each position of the $r_p-\rho_{20}$ space, the median value of the physical parameter is obtained from those of galaxies within a certain separation from the position}. Constant-parameter contours are drawn.}\label{fig6}
\end{figure*}

The  u - r color of a galaxy is a measure of the SFA of galaxies in the recent past. Equivalent width of H$\alpha$ line is a well calibrated standard indicator of star formation rate (Kennicutt 1998). This width can be defined as the ratio of the H$\alpha$ luminosity to the underlying stellar continuum thus  representing  a measure of the the current to past average star formation. It is therefore a model independent, directly observed proxy for specific star formation rate of a galaxy. The color gradient of a galaxy enables us to study the star formation history along the disk. {The {\it u-}band surface photometry of galaxies in our sample is noisy. For this reason we use gradient in g - i color which has been corrected for the inclination and seeing effects}. It is defined as the difference in g - i color between the central and annulus region of the galaxy in our sample. Observational studies indicate that the higher surface brightness galaxies have shallower color gradients.  
{We find that when the target galaxy is early type its median u - r color is constant, irrespective of the neighbor's morphology or separation. For late type target galaxies the u - r color is slightly larger for those with early type neighbors. On smaller separations the late type galaxies having late type neighbors show an increase in their u -r color, however, the sample size becomes so small that the increase is not statistically significant}.  In line with the constancy of u - r color, the equivalent width of H$\alpha$ line for early type galaxies is constant and indistinguishable of neighbor's morphology at separations larger than 0.1 $r_{vir,nei}$. The situation is completely different when the neighbor morphology is late type. {When any galaxy comes within the virial radius of its late type  neighbor it shows a dramatic increase in its SFA. The increase in width of H$\alpha$ line can be attributed to the interactions, however small, that are going on in these galaxies even though they reside in comparatively low density environments}. 
The gradient in g - i color, slightly increases as the galaxy approaches it neighbor. This indicates that central region of galaxies become slightly bluer compared to the outside region. Different neighbor morphology do not seem to affect this behavior significantly unlike the case of W(H$\alpha$) and u - r color.

In order to better understand the dependence of these physical parameter on different environment parameters, in Figure \ref{fig6}, we study their behavior in three dimensional space indicated by large scale background density, projected separation from the neighbor and neighbor's morphology. The u - r color for early type target galaxies changes little as we traverse the full $\rho_{20}$ - $r_p$ parameter space in the top left panels of Figure \ref{fig6}. For early type neighbor (E - e ) and late type neighbors (E - l) the color changes by 0.02 remaining constant when the target galaxy is outside virial radius of its nearest neighbor. As the galaxy enters the virial radius of the neighbor, the color decreases slightly and shows a weak dependence on $\rho_{20}$ for E - l case. The decrease in color on small scales for E - l galaxies is consistent with our findings of decrease in early type fraction on smaller scales in lower panel of figure \ref{fig2}. For late type target galaxies with early type neighbors, the u - r color increases as the galaxies come closer to each other within the virial radius. Outside the virial radius the color does not change with $r_p$ but shows a weak dependence on $\rho_{20}$. For the case of L-l, we notice a peculiar increase in color as the galaxies come closer in the comparatively denser regions. In lower density regions, however, the color remains constant as a function of separation between nearest neighbors. {We can conclude that color of early type galaxies remain constant as a function of small and large scale environment but for late type galaxies, we observe a weak but non zero dependence on underlying density field}. This observations also indicates that {u - r color alone can not completely quantify the star formation activity of a galaxy}. Three dimensional contours of $W(H\alpha)$ show an even clearer dependence of neighbor separation and morphology. The middle column in figure  \ref{fig6} shows that SFA of early type galaxies with early type neighbors in intermediate density field does not depend on either density or the neighbor separation. In the comparatively lower as well as higher density regions there is a slight decrease in star formation activity as the galaxies come closer. The situation is similar for L - e case where we notice a slight decrease in $W(H\alpha)$ in high density regions.
For the E - l case, we notice a complete independence from local density in all density environments. At separations larger than virial radius, $W(H\alpha)$ remains constant but {as the galaxy enters the virial radius of the late type neighbor $W(H\alpha)$ shows a gradual increase}. 
This situation is similar to L - l case, except the increase in equivalent width $W(H\alpha)$ is almost double at close separation to what it was at separations beyond the virial radius.  In the right column of Figure \ref{fig6}, contours represent the distribution of the median $\Delta (g-i)$   at each location of the $r_p$-$\rho_{20}$ space. The color gradient $\Delta (g-i)$ of all galaxies irrespective of neighbor morphology is independent of density  as can be seen from the horizontal contours. In case of E - e, E - l and L - e galaxies we notice a change in $\Delta (g-i)$ of about 0.01 for well separated and closeby neighbor galaxies. {When a late-type galaxy approaches another late type neighbor inside its  virial radius, its color gradient increases (center becomes relatively bluer) significantly compared to three other cases}.  

\subsection{\it Structure Parameters}
\begin{figure*}[htb]
\begin{center}
\includegraphics[height=19cm,width=19cm]{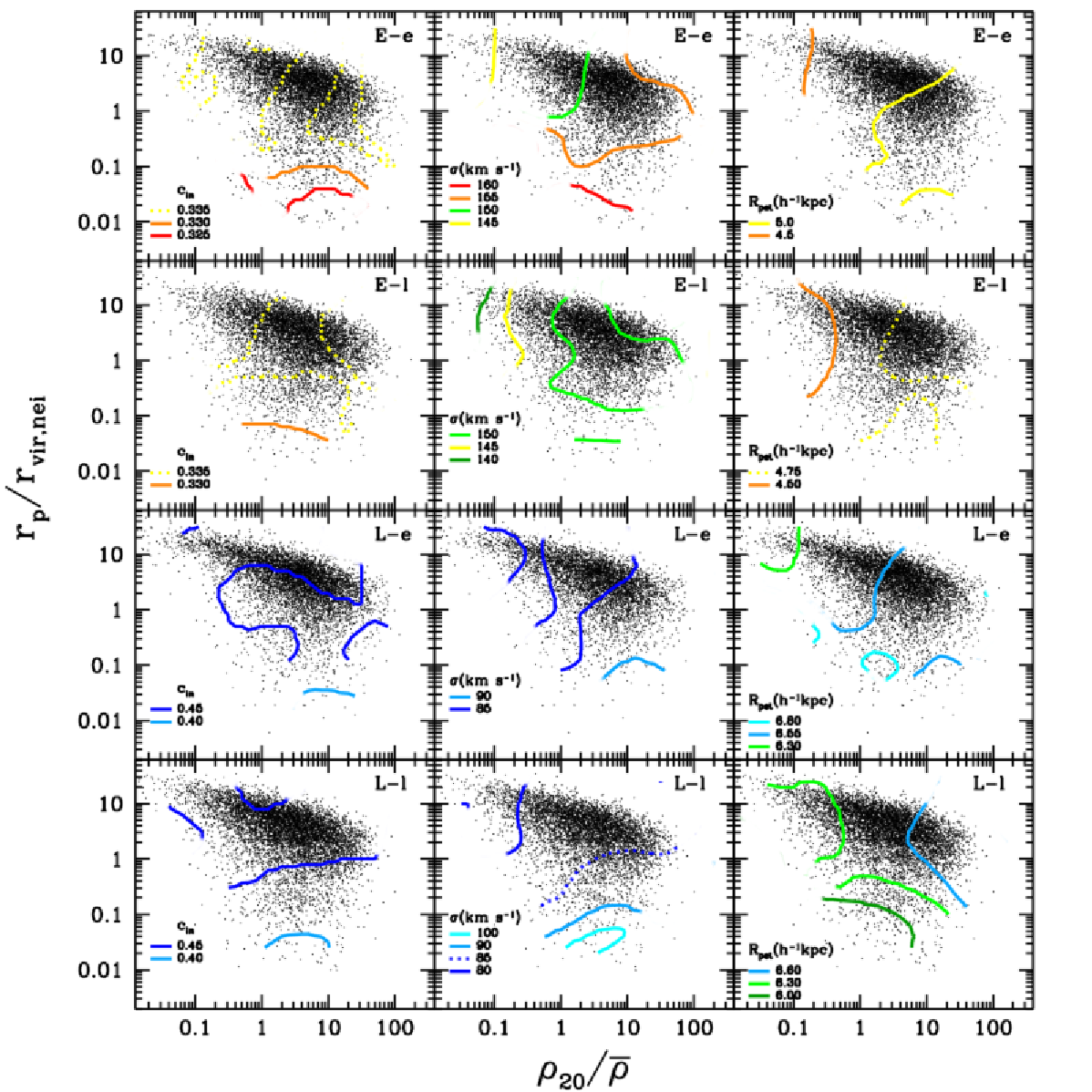}
\end{center}
\caption{Three dimensional (morphology, $r_p$ and $\rho_{20}$) environment dependence of the inverse concentration index $c_{in}$, central velocity dispersion $\sigma$ and Petrosian radius $R_{pet}$ of galaxies with $-−19.5 > Mr > -−20.5$. From top to bottom in each column, galaxies are divided into four cases: the E-e, E-l, L-e, and L-l galaxies, respectively. {The median value of the physical parameter at each location is obtained from those of galaxies within a certain spline radius from the location}. Different curves represent the constant-parameter contours}\label{fig8}
\end{figure*}
Concentration indices ($c_{in}$) provide a quantitative determination of the radial distribution of light in specified passbands that is both useful in itself, and also provides a more objective indicator of galaxy type than is provided by, {\em e.g.}, Hubble classifications (Shimasaku {\em et al.} 2001;
Strateva {\em et al.} 2001). Central velocity dispersion ($\sigma$) refers to the velocity dispersion of the interior regions of an extended object like galaxy or cluster of galaxies. Physical parameter representing size of the galaxy is the Petrosian radius ($R_{pet}$) obtained from $i-$band images of galaxies.
A three dimensional dependence of various structural parameters on morphology, large scale background density $\rho_{20}$  and projected separation from the neighbor $r_p$ is presented in  figure \ref{fig8}. The left column shows the variation of $i-$band inverse concentration ratio of galaxies.  As expected we observe a very weak decrease of inverse concentration as the target elliptical galaxy approaches its neighbor galaxy attaining a weak maximum for $E-e$ case. For late type target galaxies, however, the decrease of $c_{in}$ is comparatively stronger for decreasing $r_p$. For different combinations of target and neighbor galaxies, we find $c_{in}$ to be independent of $\rho_{20}$. The decrease in $c_{in}$ of late type galaxies at smaller neighbor separation has also been reported in Park \& Hwang
(2009) who analyzed a sample of galaxies in relatively denser environment. For early type target galaxies the concentration is almost constant as a function of separation from neighbor whereas  for late type targets it increases ($c_{in}$ decreases) as the target galaxy approaches the neighbor within its virial radius. This may be due to smaller size and compactness of the early type galaxy resulting in them being tidally more stable. The larger size and lower concentration  of late type galaxies compared to their early type counterparts makes the former tidally more vulnerable, resulting in a higher concentration of late type galaxies as they approach their neighbors within the virial radius. The significant drop in inverse concentration ratio in case of late type galaxies with late type neighbors can also result from relatively smaller  velocity difference between the target and neighbor galaxy, resulting in larger tidal energy deposits. Effect of various environments on central velocity dispersion $\sigma$ is shown in middle column of Figure \ref{fig8}. We observe a weak dependence of $\sigma$ on $\rho_{20}$ when $\rho_{20} < \bar\rho$.  For early type galaxies, $\sigma$ slightly increases with $\rho_{20}$ in the low and intermediate density regions at $r_p > r_{vir,nei}$. This behavior is consistent with similar findings using galaxies from earlier SDSS data release (Park {\em et al.} 2007; Park \& Choi
2009). However, in the high density regions we do not find any dependence of $\sigma$ on $\rho_{20}$. The E - e panel shows an increase in $\sigma$ as galaxies come closer to each other at fixed $\rho_{20}$ while in case of E - l we don't find any dependence of $\sigma$ on $r_p$. For late type target galaxies we report a $\sigma$ that is independent of $r_p$ and $\rho_{20}$ at separations larger than  $\sim$ 0.1 $r_{vir,nei}$. At closer separations $\sigma$ increases as the separation decreases particularly for L - l galaxies. This behavior can once again be explained by compact, fast and tidally more stable nature of early type galaxies. The large deposits of tidal energy in case of late-late galaxy pairs can give rise to an increase in velocity dispersion as the pairs come closer (Binney \& Tremaine 1987).
The right column of Figure \ref{fig8} shows the dependence of galaxies size on neighbor separation and background density. In case of E - e panel we notice that the size of the galaxy slightly increases as it moves in a higher density region at separation larger than the virial radius of its neighbor. Inside the virial radius we report a fixed size for E - e  galaxies. For the remaining cases also the size of galaxies does not show a significant change as the galaxy traverses the full $r_p$ - $\rho_{20}$ parameter space.  Choi {\em et al.} (2007) reported a strong dependence of galaxy size on morphology and luminosity of the galaxy population. In this work we have fixed the morphology, restricted the change in magnitude to 1 mag and omitted galaxies from high density regions of the universe resulting into the measurements of Petrosian radius that is essentially fixed across a  range of background density and neighbor separation. 
\section{Discussion and Conclusions}\label{discuss}
In this paper we have studied various  physical properties of mainly passively evolving field galaxies from low density regions of SDSS. Effects of the nearest neighbor distance,  the nearest neighbor’s morphology, and the large-scale background density are examined. In order to better understand the effect of the nearest neighbor galaxy, we have removed all those galaxies from our analysis that lie within the virial radius of their second nearest neighbor. We have hoped to remove the biases in galaxy properties due to morphology misclassification by using  an accurate automated morphology classifier combined with careful visual inspection. Mass transfer between interacting galaxies can help us better understand the reason for decrease in early type fraction of galaxies as the target galaxies approach  a late type neighbor galaxy (see Figure \ref{fig2}) within their virial radius. In such a scenario, the early type galaxy can acquire cold gas from the late type neighbor enabling the former to form a disk and transform itself to a late type galaxy (Park {\em et al.} 2008). We find that at fixed background density the isolated galaxies are comparatively brighter and among isolated galaxies the brighter ones lie in higher density environments. It can serve as the evidence of transformation of the galaxy luminosity class through the merger process.  Our analysis reconfirms the importance of radius of the galaxy plus dark halo systems as a distance scale inside which most of the galaxy properties start to be sensitive to both the nearest neighbor’s distance and its morphology.
The gravitationally bound pair of galaxies orbit each other within the virial radius resulting in repeated hydrodynamic interactions contributing  to change in properties of the orbiting galaxies. {Our study emphasize the importance of neighbor galaxy's morphology in either enhancing or suppressing of star formation activity (see e.g. middle column of Figure \ref{fig6}) of target galaxy which has long been assumed to only increase due to the internal mass perturbed by the tidal force of the neighbor (Kennicutt {\em et al.}
  1987; Nikolic {\em et al.} 2004; Patton {\em et al.} 2011)}.{ Such a relation between morphology and star formation activity can be attributed to hydrodynamic interactions between approaching galaxies. Ram pressure effects experienced due to the collision with the hot gas  of the neighbor elliptical galaxy can explain the change in star formation activity of the late type galaxy}.
\begin{equation}
W(H\alpha) = (31.37\pm5.50)exp((-0.70\pm0.46)\frac{r_p}{r_{vir,nei}}) + C
\nonumber
\end{equation}  where C is the measure of the value outside the virial radius of the neighbor.
\begin{figure}
\begin{center}
\includegraphics[height=9cm,width=8cm]{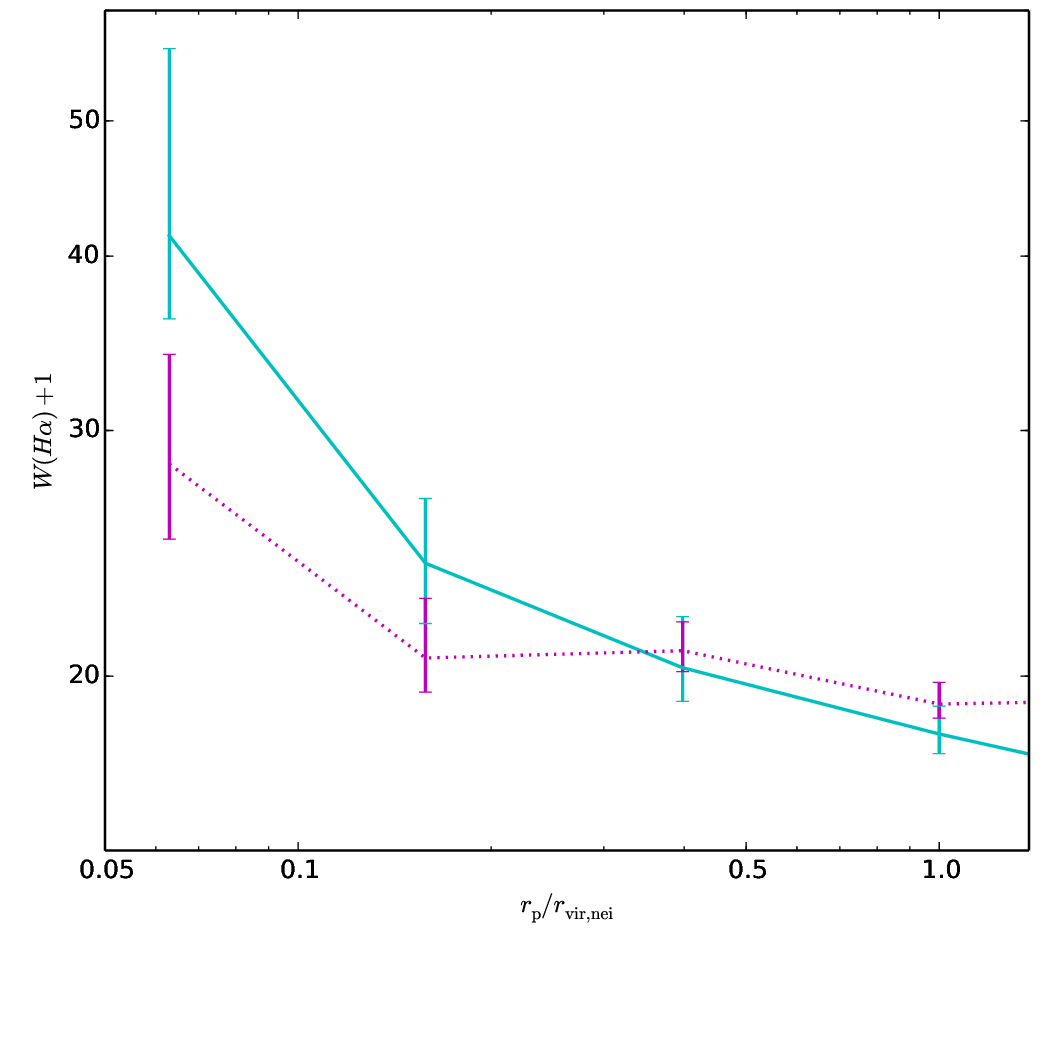}
\end{center}
\caption{Dependence of $W(H\alpha)$ - $r_p/r_{vir,nei}$ relation on the absolute magnitude of the neighbor galaxy for late type galaxies with late type neighbors. The solid {\it cyan } line is for the target galaxies having neighbors that are up to one magnitude fainter ($\Delta M_r = 1$), and dotted {\it magenta} line is for galaxies having neighbors that are  up to one magnitude brighter ($\Delta M_r = -1$) than the target itself. The $r_p$-space is uniformly binned in the logarithmic scale and in each bin, late type galaxies with axis ratio of $b/a \ge 0.6$ are selected for the median $W(H\alpha)$ curve.}\label{fig9}
\end{figure}

In order to analyze the robustness of our findings against the choice of neighbor selection parameter($\Delta M_r$), we restudied the behavior of equivalent width of $H\alpha$ line using different samples of galaxies that are constrained to have  brighter as well as fainter neighbors than themselves. Figure \ref{fig9} represents a  scenarios in which the two distinct target galaxy populations exhibit similar behavior of $W(H\alpha)$ as a function of neighbor separation. Well inside the virial radius, $W(H\alpha)$ of late-type galaxies having late-type neighbors, decreases slightly as we traverse the galaxy populations having fainter neighbors (solid cyan line, Figure \ref{fig9}) to those having brighter neighbors(dotted magenta line). We can fit the $W(H\alpha)$  as a function of $r_p/r_{vir,nei}$ using the function  
{The slight decrease in $W(H\alpha)$ as a function of neighbor brightness for passive galaxies can probably be explained in terms of exchange of gas between interacting galaxies}.

The additional role played by gravitational effects in the evolution of galaxies is evident by morphology independent structural parameter changes as a function of neighbor separation. A comparatively larger variation in $c_{in}$ and $\
\sigma$ for galaxy pairs that have smaller relative velocity suggests an inverse correlation between tidal energy deposits in a galaxy and \
velocity difference between pairs (Barton {\em et al.} 2000; Park \& Choi 2009). {The overall small variation in structural properties as a function of environments as evident in Figure \ref{fig8} suggests these are less closely related to their ``  environments'' than are their masses and star formation histories (see e.g. Blanton {\em et al.} 2005b)}. 

Major outcomes of our studies are as follows.
\begin{itemize}
\item The properties of passive galaxies that are separated more than the one virial radius of the nearest neighbor are independent of the separation. Initial dependence of physical property on the separation starts as soon as the galaxy enters the neighbor's virial radius with another change occurring at about 0.1 $r_{vir,nei}$, which corresponds to $20-30h^{-1}$kpc for galaxies in our sample.

\item We do not observe a morphology density relation for passive galaxies in low density environments as indicated by nearly horizontal, constant $f_E$ contours in Figure \ref{fig2}. The galaxy morphology mainly depends on the pair separation combined with morphology of the neighbor galaxy.

\item Only luminosity of the passive galaxies show a dependence on $\rho_{20}$ (Figure \ref{fig4}) especially at separations larger than virial radius of the neighbor. This corroborates the evidence regarding transformation of luminosities of passive galaxies through mergers.

\item At fixed morphology and luminosity, both SF activity as well as structural parameter of galaxies show a weak or negligible dependence on $\rho_{20}$. Variations in neighbor absolute magnitude  does not vary the star formation activity of the target galaxy considerably (Figure \ref{fig9}). The weak residual dependence on $\rho_{20}$ of the $u - r$ color of galaxies can be attributed to the presence of hotter and denser halo gas in galaxies which might still be in high density environments.   

\end{itemize}

\section*{Acknowledgments}
JKY thanks Changbom Park, Yun-Young Choi and Jasjeet Bagla for many useful discussions and suggestions. JKY acknowledges financial support from Chinese Academy of Sciences in the form of fellowship during this work. JKY and XC acknowledge support from NSFC via project number 11250110510 and 11373030 respectively. We thank Korea Institute for Advanced Study for providing computing resources (KIAS linux cluster system) for this work. Funding for the SDSS and SDSS-II was provided by the Alfred P. Sloan Foundation, the Participating Institutions, the National Science Foundation, the U.S. Department of Energy, the National Aeronautics and Space Administration, the Japanese Monbukagakusho, the Max Planck Society, and the Higher Education Funding Council for England. The SDSS was managed by the Astrophysical Research Consortium for the Participating Institutions.

\label{lastpage}
\end{document}